\documentclass[prl,twocolumn,showkeys,showpacs,superscriptaddress]{revtex4}
\usepackage{epsfig,latexsym}

\begin{document}


\title{Vortex Line Ordering in the Driven Three-Dimensional Vortex Glass}

\author{Ajay Kumar Ghosh}
\affiliation{Department of Physics, Ume{\aa} University, 901 87 Ume{\aa}, Sweden}
\affiliation{Department of Physics, Jadavpur University, Kolkata 700032, India}
\author{Peter Olsson}
\affiliation{Department of Physics, Ume{\aa} University, 901 87 Ume{\aa}, Sweden}
\author{S. Teitel}
\affiliation{Department of Physics and Astronomy, University of
Rochester, Rochester, NY 14627}
\date{\today}

\begin{abstract}
Resistively-shunted-junction dynamics is applied to the three dimensional uniformly frustrated XY model with randomly perturbed couplings, as a model for driven steady states in a type-II superconductor with quenched point pinning. For a disorder strength $p$ strong enough to produce a vortex glass in equilibrium, we map the phase diagram as a function of temperature $T$ and uniform driving current $I$.  Using finite size analysis and averaging over many realizations of quenched randomness we find
a first-order melting $T_m(I)$ from a vortex line smectic to an anisotropic liquid.
\end{abstract}
\pacs{74.25.Dw, 74.25.Qt, 74.40.+k, 64.60.-i}
\maketitle

Ordering and phase transitions in driven steady states far from equilibrium
remains a topic of considerable general interest. In particular, the
spatial ordering of driven vortex lines in a type-II
superconductor with random point pinning has received considerable 
theoretical \cite{Koshelev,Giamarchi,Balents,Scheidl} and experimental \cite{RDexpt} attention.
Originally, Koshelev and Vinokur proposed \cite{Koshelev} that a moving steady state would
average over quenched randomness, and that a system which was disordered in equilibrium
could reform into an ordered vortex lattice when driven.
Later, Giamarchi and Le Doussal \cite{Giamarchi} argued that this state  would
be a ``moving Bragg glass,"  with algebraically decaying
translational correlations both parallel and transverse to the driving force.  
Balents, Marchetti and Radzihovsky \cite{Balents} then argued that
the moving Bragg glass would be unstable to  dislocations which
decouple the planes of vortex lines moving parallel to the drive, resulting in  a smectic ordering. Other theoretical works \cite{Scheidl} have supported one or more
of these scenarios.

While numerous simulations have studied this problem for point vortices in two dimensions \cite{R2D}, few works have treated three dimensional (3D) vortex lines
at finite temperature  \cite{Zimanyi1, Dominguez,Chen,Nie}.
We present here the first 3D simulations to include both a systematic study of
finite size effects, as well as averaging over many realizations of the quenched
randomness \cite{note}.  Such considerations are necessary for an unambiguous
determination of ordering in driven steady states.  We map the phase diagram as
a function of temperature $T$ and driving current $I$ for the strongly random case,
and study the nature of ordering just below and above the melting transition $T_m(I)$.

Following Refs. \cite{Dominguez,Chen,Nie} we use the 3D XY model \cite{XY}
with resistively shunted junction (RSJ) dynamics \cite {Kim} to model our system.  The Hamiltonian is given by,

\begin{equation}
{\cal H}[\theta({\bf r}_i)]=-\sum_{i\mu}J_{i\mu}\cos(\theta({\bf r}_i)-\theta({\bf r}_i+\hat\mu)-A_{i\mu})
\label{eH}
\end{equation}
where $\theta({\bf r}_i)$ is the phase of the superconducting wavefunction on site ${\bf r}_i$
of a cubic $L_x\times L_y\times L_z $ grid of sites with bonds in directions $\mu=x,y,z$.  The circulation of $A_{i\mu}$
around any plaquette of the grid is fixed and equal to $2\pi f$ with $f$ the density of applied magnetic flux quanta through that plaquette.  We use a uniform value $f=1/12$ oriented in the $\hat z$ direction. The resulting field induced vortex line density, also equal to $f$, forms a vortex lattice in the 
equilibrium ground state of the pure (disorder-free) system, with basis
vectors ${\bf a}_1=4\hat y$ and ${\bf a}_2=3\hat x+2\hat y$.  To model quenched
point randomness we use couplings
\cite {OlssonMelt} $J_{i\mu}=J_\mu(1+p\epsilon_{i\mu})$,
where $\epsilon_{i\mu}$ are uncorrelated, uniformly distributed, random
variables with $\langle\epsilon_{i\mu}\rangle=0$, $\langle\epsilon_{i\mu}^2\rangle=1$. We use 
$J_\mu=J_\perp$ in the $xy$ plane, and $J_z=J_\perp/40$
to enhance vortex line fluctuations along the $\hat z$ direction.  The
disorder strength is controlled by the parameter $p$.  In our earlier work \cite{OlssonMelt,OlssonVG} we showed that above a critical $p_c$  the
vortex line lattice becomes unstable to a vortex glass at low temperatures in equilibrium.
Here we consider the strongly random limit $p=0.15>p_c\approx 0.14$.  

We use RSJ dynamics, with equation of motion as in \cite{Chen}, however with
fluctuating twist boundary conditions \cite{Kim} in {\it all} directions.
We apply a uniform current ${\bf I}=I\hat x$,  resulting in a force $\hat z\times {\bf I}$ on the vortex lines driving them in the $\hat y$ direction.
We use a second order Runge-Kutta integration method with
dimensionless time step \cite{Chen} $\Delta t=0.1$, and typically 
$2.6\times 10^6$ time steps per simulation run, resulting in a net displacement of the vortex
line center of mass of roughly $48,000$ grid spacings in the ordered phase.
Depending on system size, 
up to $3/4$ of these steps may be discarded for equilibration.  When probing 
behavior at a specific point in the $T-I$ plane, we 
usually take the pure system ground state as our initial configuration.
For numerous test cases, however, we have 
started with a random initial state at high $T$ and slowly cooled to the desired point.
Except for a narrow region of hysteresis at the melting transition, we always find the same long time steady state for both initial conditions.

To determine the structural order of our system we use the vortex structure function,
%
%
\begin{equation}
S({\bf  k})={1\over fL_xL_yL_z}\sum_{{\bf R},{\bf r}}{\rm e}^{i{\bf k}\cdot{\bf r}}
\langle n_z({\bf R}+{\bf r})n_z({\bf R})\rangle
\label{eS}
\end{equation}
where $n_z({\bf r})$ is the vortex line density in the $\hat z$ direction at position ${\bf r}$.  
We use $\langle\dots\rangle$ to denote averages over
simulation time and $[\dots ]$ to denote  averages over
independent realizations of quenched randomness (typically $40$ realizations are used).  We also consider the correlations,
\begin{eqnarray}
C(x,y,z)&=&{1\over L_xL_yL_z}\sum_{\bf k}S({\bf k}){\rm e}^{-i{\bf k}\cdot{\bf r}}\label{eC1}\\
\tilde C(x,k_y,z)&=&{1\over L_xL_z}\sum_{k_xk_z}S({\bf k}){\rm e}^{-i(k_xx+k_zz)}
\label{eC2}
\end{eqnarray}

\begin{figure}
\epsfxsize=8.6truecm
\epsfbox{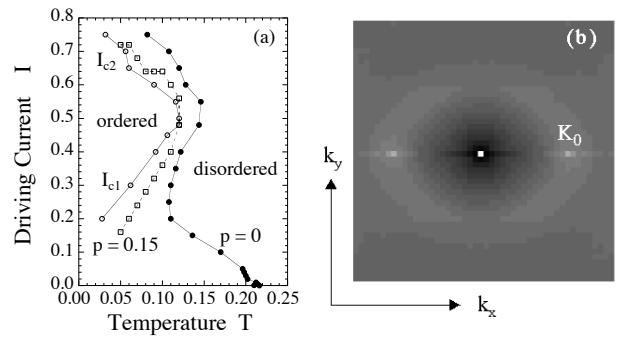}
\caption{(a) Phase diagram as  function of temperature $T$ and uniform driving current $I\hat x$ (vortex lines move in $\hat y$ direction).  Solid symbols are for a pure system, $p=0$. Open symbols are for a system with quenched point pinning, $p=0.15$;  $\circ$ is phase boundary obtained by cooling $T$ from the disordered phase at fixed $I$, while $\Box$ is phase boundary obtained by increasing or decreasing  $I$ from the ordered phase at fixed $T$.  Results are
from a $24\times 24\times 16$ size system.
(b) Intensity plot of $\ln S({\bf k}_\perp,k_z=0)$, for one particular random realization ($p=0.15$), in disordered phase at $I=0.48$, $T=0.13$ for a $48\times 48\times 48$ size system; ${\bf k}=0$ is at the center of the figure.  }
\label{f1}
\end{figure}

Using the appearance of sharp peaks in $S({\bf k}_\perp, k_z=0)$
to signal an ordered phase, in Fig.~\ref{f1}a we present the $T-I$ phase 
diagram for a $24\times 24\times 16$ sized system for both the
pure ($p=0$) and random ($p=0.15$) cases.  We measure $T$ in units of
$J_\perp$, and $I$ in units of $I_0=2e J_\perp/\hbar$. Crossing the phase boundary
at any value of the current
we find a discrete jump in energy, suggesting a first order melting transition $T_m(I)$.
For the random case our
phase boundary is from a single random realization only and is determined
by two separate methods: by cooling in $T$ at fixed $I$ 
from the disordered phase (solid line); by increasing and decreasing 
$I$ at fixed $T$ from within the ordered phase (dashed line). Assuming some hysteresis, 
as in a first order transition, the two methods give reasonable
agreement.  At fixed $T$, the random system orders only for a finite current interval
$I_{c1}(T)<I< I_{c2}(T)$.  For the pure case,  our $T_m(I)$ looks 
qualitatively similar to that found in Ref.[\onlinecite{Chen}] for the 
weakly random case $p<p_c$.  The melting of the ordered phase upon increasing $I>I_{c2}$,
which exists for both random and pure systems,
is contrary to theoretical predictions \cite{Koshelev,Giamarchi,Balents,Scheidl}.
Ref.~[\onlinecite{Chen}] suggested this to be a consequence of thermally excited vortex loops.
We in fact find that the density of  thermally excited loops is negligible everywhere 
in the ordered phase, but shows a finite jump to a noticeable amount as $T_m(I)$
is crossed; this jump is particularly large as $I$ crosses $I_{c2}(T)$.

\begin{figure}
\epsfxsize=8.6truecm
\epsfbox{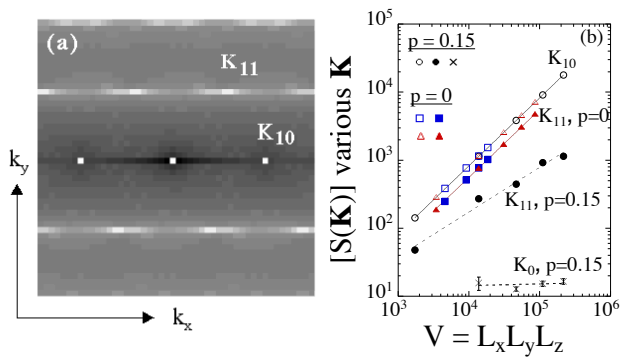} 
\caption{(a) Intensity plot of $\ln S({\bf k}_\perp,k_z=0)$, for one particular random realization, in ordered phase at $I=0.48$, $T=0.09$ for a $L_x=L_y=L_z=48$ size system; ${\bf k}=0$ is at the center of the figure.
(b) Scaling of disorder averaged peaks heights vs. system volume $V=L_xL_yL_z$.  
For the ordered state of (a) we plot
$[S({\bf K}_{10})]$ and $[S({\bf K}_{11})]$ for pure ($p=0$) systems of size $24\times 24\times L$ (squares) and $L\times L\times 24$ (triangles), and for systems with quenched point pinning ($p=0.15$) of size $L^3$, $L=12, 24, 36, 48, 60$ (circles).  We
also plot $[S({\bf K}_0)]$ for the $p=0.15$ disordered state at $T=0.13$ of Fig.~\protect\ref{f1}b for systems of size $L^3$ with $L=24,36,48,60$ (crosses).
}
\label{f2}
\end{figure}

Consider first the disordered phase. Fig.~\ref{f1}b shows an intensity plot of
$\ln S({\bf k}_\perp,k_z=0)$ for a single realization of a random
 $48^3$ system at $I=0.48$, $T=0.13$, just above the maximum
in $T_m(I)$. Two relatively faint peaks, $S({\bf K}_0)/S(0)\simeq 0.008$, 
lie off the origin along the $k_x$ axis.  Previous
works \cite{Chen,Nie} have interpreted such peaks as evidence for a smectic phase.
However, as shown in Fig.~\ref{f2}b, 
we find no significant increase of peak height $S({\bf K}_0)$ with system size, thus indicating that this is
an anisotropic liquid with only short ranged translational correlations.

Next consider the ordered phase.  Fig.~\ref{f2}a shows an intensity plot
of $\ln S({\bf k}_\perp, k_z=0)$ for a single realization of a random
$48^3$ system at $I=0.48$, $T=0.09$, just below the maximum
in $T_m(I)$.  Now sharp peaks, $S_{\rm peak}/S(0)\simeq 1$, lie along
the $k_x$ axis.
In the {\it pure} system, the vortex lattice state has Bragg peaks at 
${\bf K}_{10}=(2\pi/3)\hat x$ and ${\bf K}_{11}
=(2\pi)(\hat x/6+\hat y/4)$, as labeled in Fig.~\ref{f2}a.  In the random system 
a peak remains exactly at ${\bf K}_{10}$; a second peak is always found with
$K_y=K_{11,y}$ but with $K_x-K_{11,x}=0,\pm 2\pi/L_y$ depending on the particular random
realization.  We therefore generalize our notation to denote by ${\bf K}_{11}$ the exact location
of this second peak, with the understanding that the value of $K_{11,x}$ may shift
slightly between different random realizations.  In Fig.~\ref{f2}b we plot the disorder
averaged height of these peaks, $[S({\bf K}_{10})]$ and $[S({\bf K}_{11})]$, vs.
system volume $V=L_xL_yL_z$, for several different system sizes, 
for both the pure and random cases. 
In both cases $[S({\bf K}_{10})]$ scales linearly with $V$, indicating a sharp Bragg
peak. This Bragg peak indicates that vortex lines are organized into specific $yz$ planes
with a periodic spacing of $3\hat x$ between planes.  Ordering within and between planes
is reflected in the scaling of $[S({\bf K}_{11})]$.  For the pure case, 
$S({\bf  K}_{11})\sim V$ indicating the long range translational order of a
moving vortex line lattice.  For the random case, however, $[S({\bf K}_{11})]$
grows less rapidly than $V$.
The dashed line in Fig.~\ref{f2}b represents a power law divergence of $[S({\bf K}_{11})]\sim V^{2/3}\sim L^2$, however if we discard the smallest point at $L=12$, the exponent of a power law fit in $L$ becomes $1.65$.  It is thus unclear from Fig.~\ref{f2}b whether our
biggest size system has reached the asymptotically large $L$ limit.  
The sublinear growth of
$[S({\bf K}_{11})]$ with $V$ indicates that the system has less than long range translational 
order.  However it does not indicate which direction(s) are less ordered than others.  For this 
we will consider the shape of the peak at ${\bf K}_{11}$ and the
real space correlations of Eqs.~(\ref{eC1}) and (\ref{eC2}).

\begin{figure}
\epsfxsize=8.6truecm
\epsfbox{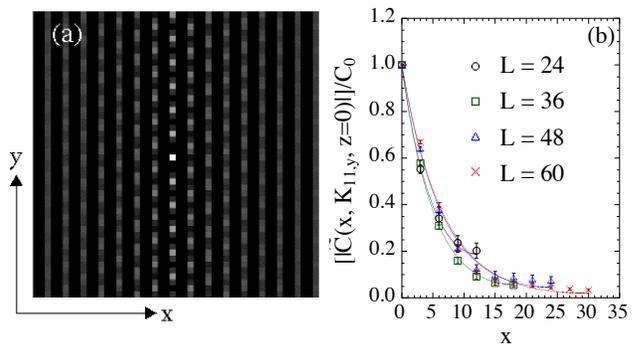}
\caption{(a) Intensity plot of real space correlation, $C(x,y,z=0)$, for one particular random
realization, in ordered phase at $I=0.48$, $T=0.09$, $p=0.15$, for a $L_x=L_y=L_z=48$ size system; ${\bf r}=0$ is at the center of the figure.
(b) Corresponding disorder averaged correlation, $[|\tilde C(x,K_{11,y},z=0)|]$ vs. $x$ for systems of size $L\times L\times L$.
}
\label{f3}
\end{figure}

The sharpness of the peak at ${\bf K}_{11}$ in the $\hat k_y$ direction
suggests that vortex lines are periodically ordered along their direction of motion $\hat y$
within each $yz$ plane.  That this peak appears broad in the $\hat k_x$ direction
suggests that the $yz$ planes  have only short range 
correlations between them.  We thus refer to the $yz$ planes containing the vortex lines as {\it smectic} planes.
We now consider the real space correlations
of Eqs.~(\ref{eC1}) and (\ref{eC2}).  Fig.~\ref{f3}a shows an intensity plot of
$C(x,y,z=0)$ corresponding to $S({\bf k})$ of Fig.~\ref{f2}a.  We clearly see the 
structure proposed above: vortices lie in periodically spaced $yz$ planes with 
separation $3\hat x$; within a given plane, i.e. $x=0$, vortices are periodic with
separation $4\hat y$; at finite transverse separation, i.e.  $|x|>0$, the variation of $C(x,y,0)$ with
$y$ decreases, until at large $|x|$ it is almost uniform in $y$, thus indicating
short ranged correlations between the smectic planes.  To quantify this, we consider
the correlation Eq.~(\ref{eC2}) evaluated at $k_y=K_{11,y}$
giving the periodicity within a given smectic plane.  Since the slightly different
values of $K_{11,x}$ for the different random realizations give rise to different complex
phase shifts in $\tilde C(x,K_{11,y},z=0)$, we disorder average the absolute value.
In Fig.~\ref{f3}b we plot $[|\tilde C(x,K_{11,y},0)|]/C_0$ vs. $x$, for different system sizes $L^3$.  We normalize by $C_0\equiv [|\tilde C(0,K_{11,y},0)|]$
to better compare different system sizes.
We see a clear exponential decay to zero; solid lines are fits to a periodic
exponential giving correlation lengths in the range $\xi_x\sim 5-6.5$.  
In recent works \cite{Chen,Nie} pictures similar to Fig.~\ref{f2}a were
identified as a ``moving Bragg glass."  However the short range correlations
of Fig.~\ref{f3}b clearly show our ordered phase to be a smectic rather than a Bragg glass.

\begin{figure}
\epsfxsize=8.6truecm
\epsfbox{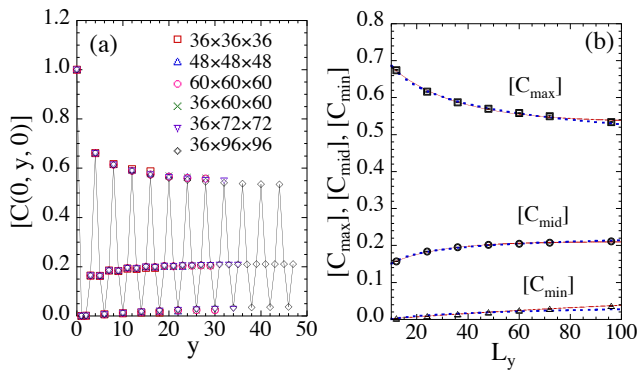}
\caption{(a) Disorder averaged real space correlation, $[C(x=0,y,z=0)]$ vs. $y$ for various system sizes, in the ordered phase at $I=0.48$, $T=0.09$, $p=0.15$. 
(b) Corresponding limiting values of maximum, middle, and minimum envelop of $[C(x=0,y,z=0)]$
(see text) vs. $L_y$ for various system sizes; red solid line is a fit to an exponential decay to a constant, blue dashed line is a fit to an algebraic decay to $1/4$.
}
\label{f4}
\end{figure}

Next we consider correlations parallel to the direction of motion $\hat y$, within
a given smectic plane.  Fig.~\ref{f4}a shows $[C(0,y,0)]$ vs. $y$, for various
system sizes, for the same $I=0.48$, $T=0.09$ as in Figs.~\ref{f2} and \ref{f3}.
Since we found $\xi_x\sim 6$ is fairly small, we include also systems
of size $36\times L\times L$ in order to obtain larger values of $L_y$.  
Fig.~\ref{f4} suggests decay to a periodic oscillation, with only a small finite 
size effect when $y\sim L_y/2$.
The large $y$ limit can be described by the three envelop
functions, 
$[C_{\rm max}(L_y)]\equiv {\rm min}_{m} [C(0,4m,0)]$, 
$[C_{\rm min}(L_y)]\equiv{\rm max}_{m} [C(0,4m+2,0)]$, and 
$[C_{\rm mid}(L_y)]\equiv{\rm max}_{m} [C(0,4m\pm 1,0)]$, 
for $m$ integer, which we plot vs. $L_y$ in 
Fig.~\ref{f4}b.  For long range translational order, these
should converge to different constants as $L_y\to \infty$.  For algebraic or
short range order, these should all converge to the value $1/4$ characterizing a uniform
vortex line density.  We find that
both exponential decay to long range order (red solid line),
and an algebraic decay to $1/4$ (blue dashed line), give equally good fits.  
For long range order, the fit to $[C_{\rm max}]$
gives a decay length $\xi_y\sim 25$, while the algebraic fit  to $[C_{\rm max}]$ gives 
a power law exponent of $\sim 0.2$.

\begin{figure}
\epsfxsize=8.6truecm
\epsfbox{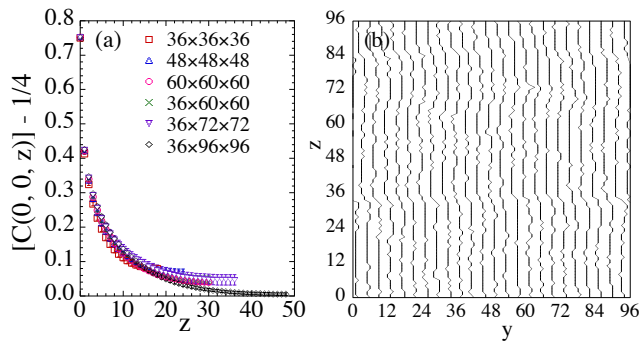}
\caption{(a) Disorder averaged real space correlation, $[C(x=0,y=0,z)]-1/4$ vs. $z$ for various
system sizes, in the ordered phase at $I=0.48$, $T=0.09$, $p=0.15$.  (b) Instantaneous vortex line configuration in a particular smectic plane of a particular random realization, in the
ordered phase at the same parameters as (a), for a system of size $36\times 96\times 96$.
}
\label{f5}
\end{figure}

Finally we consider the correlations along $\hat z$, parallel to the applied magnetic field,
within a given smectic plane.  Fig.~\ref{f5}a shows $[C(0,0,z)]-1/4$ vs. $z$,
for various system sizes, for the same $I=0.48$, $T=0.09$ as before.  The subtracted
value $1/4$ is the uniform vortex density in the smectic plane, in the absence of any ordering.
For the largest system size $L_y=96$ we find an exponential decay to zero with a
decay length $\xi_z\sim 9$.  Smaller sizes show a decay to a small finite constant that decreases
with increasing $L_y$, consistent with our earlier observation in Fig.~\ref{f2}b that
$[S({\bf K}_{11})]$ has not yet reached the asymptotic large $L$ limit for $L\leq 60$.
In Fig.~\ref{f5}b we show an instantaneous
configuration of vortex lines in a particular smectic plane.
At fixed $z$, the lines appear periodically spaced along $y$, in agreement with
Fig.~\ref{f4}.  However, tracing any particular line along $z$, we see that
it wanders an amount comparable to the inter-line spacing, consistent with the decay seen
in Fig.~\ref{f5}a.  
In some planes, such as the one shown in Fig.~\ref{f5}b, lines can have a net
tilt, by closing onto one of their neighbors under the periodic boundary condition
along $\hat z$.
Each such tilted plane must be compensated by another plane that tilts in the
opposite direction, so that the net vorticity in the $\hat y$ direction vanishes.
Looking at individual configurations, we find a strong relation between the correlations
along $\hat z$ within a given smectic plane, and the correlations along $\hat x$ between 
different smectic planes. When the lines in a smectic plane are tilted, 
as in Fig.~\ref{f5}b, or otherwise
have a transverse wandering comparable to the average spacing between lines, we find that
this plane decouples from its neighbors, moving either diffusively with respect to
its neighbors, or even with a slightly different average speed.  Planes whose lines have
small transverse wanderings, remain strongly correlated with their neighbors.  It is
necessary to have only a few such decoupled smectic planes in order to have short
range average correlations along $\hat x$, as in Fig.~\ref{f3}b.

Our analysis of the ordered phase has been for a point just below the peak in the
first order melting curve $T_m(I)$.  Establishing the nature of correlations throughout
the ordered phase would require similar finite-size, random-realization-averaged,
analyses elsewhere in this region.  Our preliminary investigations suggest that
as either $T$ or $I$ is decreased, finite correlation lengths grow and become 
too large to make such studies feasible at present.  We thus cannot rule out the
possibility of a more ordered ``moving Bragg glass" phase at lower temperatures.



This work was supported by DOE grant DE-FG02-06ER46298, by
Swedish Research Council contract No. 2002-3975, and by the resources
of the Swedish High Performance Computing Center North (HPC2N).
Travel was supported by NSF grant INT-9901379.



\begin{thebibliography}{99}
\bibitem{note} Ref.~[\protect\onlinecite{Dominguez}] does finite size scaling, but no
disorder averaging.  The high vortex density $f=1/6$ used in this work may also
enhance grid effects as compared to later works.
\end{thebibliography}
\end{document}